\documentclass[useAMS,usenatbib,letterpaper]{mnras}

\usepackage{epsfig}
\usepackage{amsmath}
\usepackage{natbib}
\usepackage{aas_macros,amssymb}
\begin{document}

\title[Bias in Discrete Surveys]{The Bias of the Log Power Spectrum for Discrete Surveys}

\author[A. Repp \& I. Szapudi]{Andrew Repp\ \& Istv\'an Szapudi\\Institute for Astronomy, University of Hawaii, 2680 Woodlawn Drive, Honolulu, HI 96822, USA}

\date{MNRAS 475, L6–L10 (2018)}

\label{firstpage}
\pagerange{\pageref{firstpage}--\pageref{lastpage}}
\maketitle

\begin{abstract}
A primary goal of galaxy surveys is to tighten constraints on cosmological parameters, and the power spectrum $P(k)$ is the standard means of doing so. However, at translinear scales $P(k)$ is blind to much of these surveys' information -- information which the log density power spectrum recovers. For discrete fields (such as the galaxy density), $A^*$ denotes the statistic analogous to the log density: $A^*$ is a `sufficient statistic' in that its power spectrum (and mean) capture virtually all of a discrete survey's information. However, the power spectrum of $A^*$ is biased with respect to the corresponding log spectrum for continuous fields, and to use $P_{A^*}(k)$ to constrain the values of cosmological parameters, we require some means of predicting this bias. Here we present a prescription for doing so; for \textit{Euclid}-like surveys (with cubical cells 16$h^{-1}$ Mpc across) our bias prescription's error is less than 3 per cent. This prediction will facilitate optimal utilization of the information in future galaxy surveys.
\end{abstract}

\section{Introduction}
Cosmology seeks to determine the precise values of cosmological parameters, and this endeavor demands full utilization of the information in galaxy surveys. The power spectrum $P(k)$ of the overdensity field $\delta$ is the standard summary statistic for this purpose. However, pushing surveys to increasingly small scales does not proportionately increase the Fisher information in $P(k)$ \citep{InfoPlat}, since much of the survey's information escapes from $P(k)$ at translinear wavenumbers ($k \ga 0.1$).

To recover this information, \citet{CarronSzapudi2013} introduce the concept of sufficient statistics, namely, observables which capture all of a field's information. They show, under the assumption of Gaussian initial conditions, that the log density $A = \ln(1+\delta)$ closely approximates a sufficient statistic when the slope of the nonlinear power spectrum is close to $-1$ (as simulations show it to be, over a wide range of scales). Thus, its power spectrum $P_A(k)$ \citep{NSS09} and mean $\overline{A}$ embody essentially all of the survey's information for parameters that affect the amplitude of power spectrum values $P_A(k)$. \citet{Repp2017} provide a simple fit for $P_A(k)$, and \citet{predA} provide a related prescription for $\overline{A}$; they also show that a Generalized Extreme Value (GEV) distribution describes $A$ well. Other characterizations of the continuous density field include those of \citet{Uhlemann2016}, \citet{Shin2017}, and \citet{Klypin2017}.

However, the $A$-statistic does not directly apply to discrete fields like the galaxy density, and thus \citet{CarronSzapudi2014} investigate $A^*$, the optimal observable for discrete fields. \citet{WCS2015} show that the power spectrum $P_{A^*}(k)$ of this statistic is biased with respect to the (continuous) log spectrum $P_A(k)$. Since our ultimate purpose is to use $P_{A^*}(k)$ to constrain the values of cosmological parameters, we require a means of predicting this power spectrum for any reasonable parameter set -- and thus we require an accurate description of this bias.

This Letter presents an \textit{a priori} means of predicting this $A^*$-bias for surveys. Using multiple realizations of the Millennium Simulation dark matter distribution \citep{MillSim}, we show that our prescription is accurate (within 5--6 per cent) for pixel number densities $\overline{N} \ga $ 1. For \textit{Euclid}-like surveys, the accuracy is better than 3 per cent for cubical survey cells with side length 16$h^{-1}$ Mpc.

 The structure of this Letter is as follows: Section~\ref{sec:Astar} characterizes the sufficient statistic $A^*$, and Section~\ref{sec:prediction} provides our bias prescription; we quantify its accuracy in Section~\ref{sec:acc} and summarize in Section~\ref{sec:concl}.

\section{The Discrete Sufficient Statistic $A^*$}
\label{sec:Astar}

Galaxy counts represent a discrete sampling of the underlying dark matter distribution. Constructing an optimal observable for such fields requires knowledge of two distributions. First, one must characterize the underlying continuous probability distribution $\mathcal{P}(\delta)$ -- or, in log space, $\mathcal{P}(A)$. Second, one must specify the conditional probability of observing $N$ objects in a survey cell given an underlying value of $\delta$ (or $A$): that is, one requires an expression for $\mathcal{P}(N|A)$. The most widespread of such schemes is local Poisson sampling, where for a given dark matter log density $A$,
\begin{equation}
\label{eq:Poisson}
\mathcal{P}(N|A) = \frac{1}{N!} (\overline{N} e^A )^N \exp(-\overline{N}e^A),
\end{equation}
$\overline{N}$ being the mean number of galaxies per survey cell. Following \citet{CarronSzapudi2014}, we define $A^*$ as the value of $A$ that maximizes $\mathcal{P}(A)\mathcal{P}(N|A)$, so that for an observed number of objects $N$ in a cell, $A^*(N)$ is the Bayesian reconstruction of the dark matter log density $A$ in that survey pixel. \citet{CarronSzapudi2014} show that $A^*$ closely approximates a sufficient statistic for galaxy surveys, and thus the power spectrum $P_{A^*}(k)$ of the $A^*$-field (together with its mean $\overline{A^*}$) contains essentially all cosmological information present in the set of galaxy counts. 

\citet{CarronSzapudi2014} specifically consider the case of a lognormal matter distribution with local Poisson sampling. Then $A^*$ is the solution to the equation
\begin{equation}
e^{A^*} + \frac{A^*}{\overline{N} \sigma_A^2} = \frac{N-1/2}{\overline{N}}.
\label{eq:Astarln}
\end{equation}

Although the lognormal distribution is a good description of the projected (two-dimensional) matter distribution, for the three-dimensional field a Generalized Extreme Value (GEV) distribution \citep{predA} fits better:
\begin{equation}
\label{eq:GEV}
\mathcal{P}(A) = \frac{1}{\sigma_G} t(A)^{1+\xi} e^{-t(A)},
\end{equation}
where
\begin{equation}
\label{eq:GEV_t}
t(A) = \left(1 + \frac{A - \mu_G}{\sigma_G}\xi\right)^{-1/\xi}.
\end{equation}
Here, $\mu_G$, $\sigma_G$, and $\xi$ are, respectively, location, scale, and shape parameters which depend on the mean $\overline{A}$, variance $\sigma_A^2$, and skewness $\gamma_1$ of $A$ as follows:
\begin{equation}
\gamma_1 = -\frac{\Gamma(1-3\xi) - 3\Gamma(1-\xi)\Gamma(1-2\xi) + 2\Gamma^3(1-\xi)}{\left(\Gamma(1-2\xi) - \Gamma^2(1-\xi)\right)^{3/2}}
\end{equation}
\begin{equation}
\sigma_G = \sigma_A \xi \cdot \left(\Gamma(1-2\xi) - \Gamma^2(1-\xi)\right)^{-1/2}
\end{equation}
\begin{equation}
\mu_G = \overline{A} - \sigma_G \frac{\Gamma(1-\xi) - 1}{\xi}
\end{equation}

For Poisson sampling of a GEV distribution, we take Equations~\ref{eq:Poisson} and \ref{eq:GEV} as our $\mathcal{P}(N|A)$ and $\mathcal{P}(A)$. To calculate $A^*$ we can thus maximize the expression
\begin{equation}
-\ln \sigma_G + (1+\xi) \ln t(A) - t(A) - \ln N! - \overline{N}e^A + N(\ln \overline{N} + A);
\end{equation}
requiring the derivative with respect to $A$ to vanish, we obtain the following equation for $A^*$ in the GEV case:
\begin{equation}
\label{eq:AstarGEV}
\frac{1}{\sigma_G} \left( 1 + \frac{A^* - \mu_G}{\sigma_G} \xi \right)^{-1-\frac{1}{\xi}} + N = \frac{1+\xi}{\sigma_G + (A^* - \mu_G)\xi} + \overline{N}e^{A^*}.
\end{equation}
It is Equation~\ref{eq:AstarGEV} which we use, together with the fits in \citet{predA}, to calculate $A^*$ throughout this Letter (except for the small squares in Figs.~\ref{fig:C_and_bias} and \ref{fig:Nbar_plot}, which use Equation~\ref{eq:Astarln}). However, we emphasize that the bias formulas we derive in Section~\ref{sec:prediction} are independent of the particular choice of $\mathcal{P}(A)$ or $\mathcal{P}(N|A)$.

\begin{figure}
\leavevmode\epsfxsize=9.5cm\epsfbox{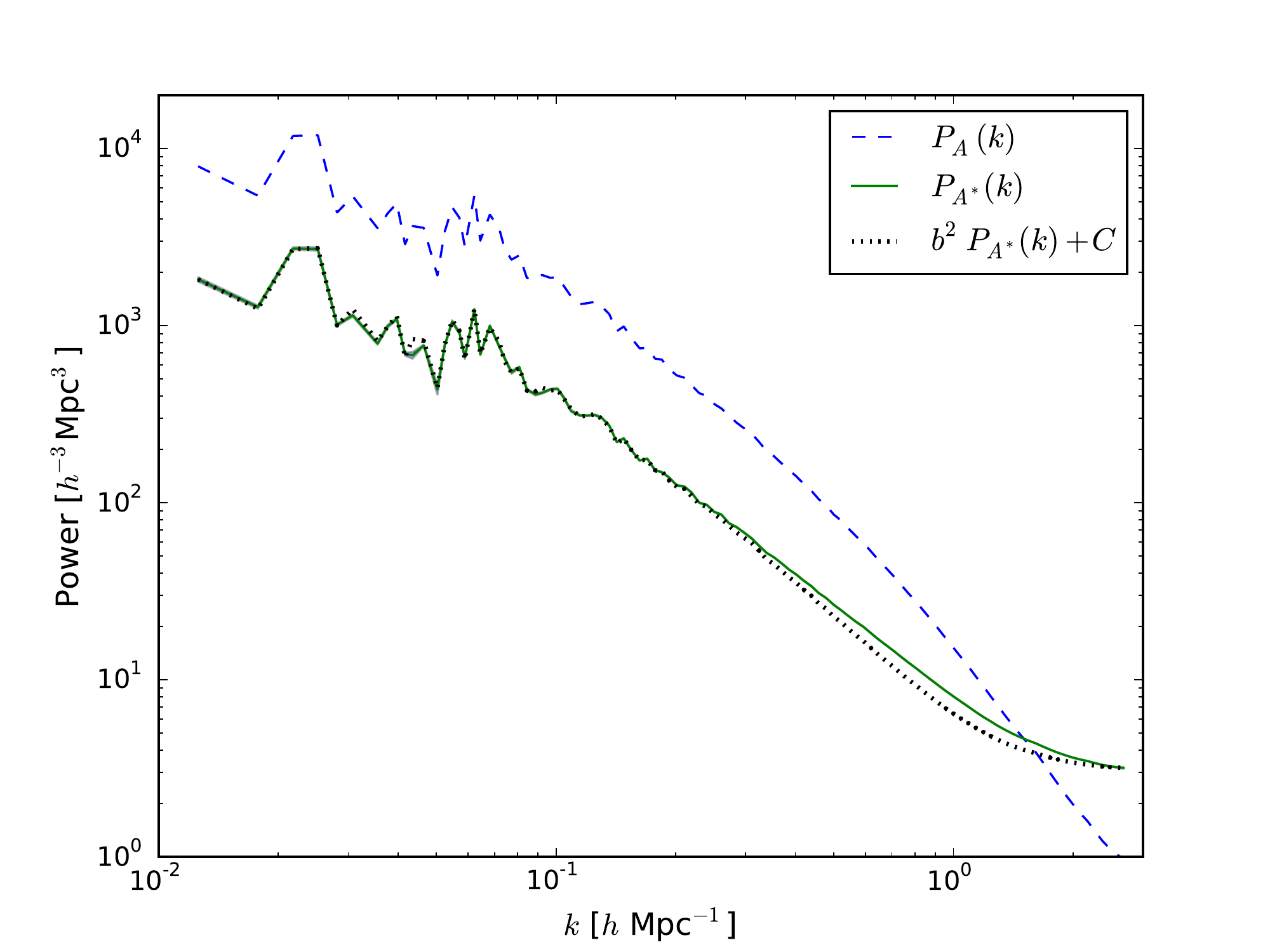}
\caption{The power spectra of $A$ (dashed) and of $A^*$ (solid) for a Poisson sampling ($\overline{N}=0.5$) of the Millennium Simulation ($z=0$). The overall bias (amplitude shift) of $P_{A^*}$ is evident, as is the high-$k$ discreteness flattening. The dotted curve shows that simply applying a bias and an additive constant to $P_A(k)$ does not fully reproduce the shape of $P_{A^*}(k)$ beyond $k \sim 0.2$.}
\label{fig:Astar}
\end{figure}

Since our ultimate purpose is to use the power spectrum of $A^*$ to constrain the values of cosmological parameters, we require a means of predicting this power spectrum for any reasonable parameter set. \citet{Repp2017} show how to predict $P_A(k)$, the power spectrum  of the continuous log density field, to an accuracy of a few per cent\footnote{We note that this prediction contains several phenomenological parameters and has to date been tested only on the Millennium Simulation and its rescalings.}. However, Fig.~\ref{fig:Astar} shows that passage from $P_A(k)$ to $P_{A^*}(k)$ introduces three effects. The most salient of these is the bias of $P_{A^*}(k)$ as a whole \citep{WCS2015}, evident in the figure as a vertical shift. Also apparent is a discreteness plateau at high wavenumbers, analogous to the $1/\overline{n}$ shot noise term in $P(k)$ (though the analogy is inexact due to the nonlinearity of the log- and $A^*$-transforms). Finally, there is a slight change of shape at intermediate wavenumbers, such that multiplying $P_A(k)$ by the bias and then adding a constant does not completely reproduce the shape of $P_{A^*}(k)$.

The bias is the factor most important for parameter constraint and is thus the focus of this letter. Hence, ignoring high-$k$ effects, we write
\begin{equation}
P_{A^*}(k) = b_{A^*}^2 P_A(k).
\label{eq:biasdef}
\end{equation}
(By convention, the ratio of the power spectra yields the square of the relevant bias factor.) Note also that \citet{WCS2015} use $b_{A^*}^2$ to refer to the reciprocal of the quantity so denoted in Equation~\ref{eq:biasdef}.

\section{Predicting the Bias of $A^*$}
\label{sec:prediction}
Passing from Fourier to real space, Equation~\ref{eq:biasdef} implies that
\begin{equation}
\xi_{A^*}(r) = b_{A^*}^2 \xi_A(r),
\end{equation}
where $\xi$ is the two-point correlation function. Now let $f(A_1, A_2)$ be the joint probability distribution function for two points separated by a distance $r$. Then
\begin{equation}
\label{eq:xiA}
\xi_A(r) = \int dA_1\,dA_2\, (A_1 - \overline{A})(A_2 - \overline{A}) f(A_1, A_2),
\end{equation}
where $\overline{A}$ denotes the mean value of $A$.
To write the analogous expression for $A^*$, we begin with the joint $A^*$-distribution:
\begin{equation}
\label{eq:jntdist}
f_{A^*}(A^*_1, A^*_2) = \int dA_1\,dA_2\,f(A_1, A_2) \mathcal{P}(A^*_1 | A_1) \mathcal{P}(A^*_2 | A_2).
\end{equation}
Given a distribution $\mathcal{P}(A)$ and a mean number density $\overline{N}$, every natural number $N$ corresponds to a discrete value $A^*(N)$ obtainable from Equation~\ref{eq:AstarGEV}. We may thus slightly abuse the notation and, in the integrand of Equation~\ref{eq:jntdist}, write $N$ instead of the corresponding $A^*(N)$:
\begin{equation}
f_{A^*}(A^*_1, A^*_2) = \int dA_1\,dA_2\,f(A_1, A_2) \mathcal{P}(N_1 | A_1) \mathcal{P}(N_2 | A_2)\\,
\end{equation}
where $A^*_1 = A^*(N_1)$ and $A^*_2 = A^*(N_2)$. Therefore the expression analogous to Equation~\ref{eq:xiA} (summing over the discrete variable rather than integrating) is
\begin{eqnarray}
\xi_{A^*}(r) & = & \sum_{N_1, N_2} (A^*_1 - \overline{A^*})(A^*_2 - \overline{A^*}) f_{A^*}(A^*_1, A^*_2) \hspace{1.1cm}\\
& = & \sum_{N_1, N_2} (A^*_1 - \overline{A^*})(A^*_2 - \overline{A^*}) \nonumber\\
& & \hspace{-.3cm}\times\int dA_1\, dA_2 f(A_1, A_2) \mathcal{P}(N_1 | A_1) \mathcal{P}(N_2 | A_2),
\label{eq:xiAstar}
\end{eqnarray}
where (as throughout this section) $A^*_1$ and $A^*_2$ are the values of $A^*$ corresponding to $N_1$ and $N_2$, respectively.

We now require an expression for $f(A_1, A_2)$. If the correlation of $A_1$ and $A_2$ is weak (as expected on large scales), we can work to first order in $\xi_A(r)/\sigma_A^2$. For a bivariate Gaussian distribution, expansion to this order yields
\begin{equation}
f(A_1,A_2) = \mathcal{P}(A_1)\mathcal{P}(A_2)\left\{1 + \frac{\xi_A(r)(A_1 - \overline{A})(A_2 - \overline{A})}{\sigma_A^4} \right\}.
\label{eq:jntprob}
\end{equation}
Equation~\ref{eq:jntprob} is our ansatz for the joint distribution of $A$ in the weak correlation limit. In other words, we assume that the distribution has the (joint) behavior of a Gaussian to first order in $\xi_A(r)/\sigma^2_A$.

To simplify notation, we temporarily define a discreteness factor that measures the fluctuations of $A^*$ given a particular value of $A$:
\begin{equation}
\langle \Delta A^* \rangle_A = \sum_N (A^* - \overline{A^*}) \mathcal{P}(N | A);
\label{eq:DA}
\end{equation}
here again we write $A^*$ for $A^*(N)$. Then Equation~\ref{eq:xiAstar} becomes
\begin{eqnarray}
\xi_{A^*}(r) & = & \int dA_1\,dA_2\, \mathcal{P}(A_1) \mathcal{P}(A_2) \langle \Delta A^* \rangle_{A_1} \langle \Delta A^* \rangle_{A_2} \hspace{0.7cm}\nonumber\\
& & \times \left\{ 1 + \frac{\xi_A(r)(A_1 - \overline{A})(A_2 - \overline{A})}{\sigma_A^4} \right\} \\
& = & \left\{ \int dA\,\mathcal{P}(A) \langle \Delta A^* \rangle_A \right\}^2 \nonumber\\
& & \hspace{0.3cm}+ \frac{\xi_A(r)}{\sigma_A^4} \left\{ \int dA\,(A-\overline{A})\mathcal{P}(A) \langle \Delta A^* \rangle_A \right\}^2.
\label{eq:twoints}
\end{eqnarray}

The first integral in Equation~\ref{eq:twoints} must vanish: by reference to Equation~\ref{eq:DA}, this integral becomes
\begin{equation}
\int dA\, \sum_N (A^* - \overline{A^*})\mathcal{P}(A) \mathcal{P}(A^* | A) = 0.
\end{equation}
Therefore
\begin{equation}
\xi_{A^*}(r) = \frac{\xi_A(r)}{\sigma_A^4} \left\{ \int dA\,(A-\overline{A})\mathcal{P}(A) \langle \Delta A^* \rangle_A \right\}^2,
\end{equation}
and expanding $\langle \Delta A^* \rangle_A$, we conclude that
\begin{equation}
b_{A^*}^2 = \frac{1}{\sigma_A^4}\left\{\sum_N \int dA\,(A-\overline{A})(A^*-\overline{A^*}) \mathcal{P}(N|A)\mathcal{P}(A) \right\}^2.
\label{eq:bAstar}
\end{equation}
Once again note that we here write $A^*$ as an abbreviation $A^*(N)$. The calculation of the mean $\overline{A^*}$ for use in Equation~\ref{eq:bAstar} is straightforward:
\begin{equation}
\overline{A^*} = \sum_N A^*(N) \int dA \, \mathcal{P}(A) \mathcal{P}(N|A)
\label{eq:Astarbar}
\end{equation}

Note also that we have written Equation~\ref{eq:bAstar} in a form conducive to calculating $b_{A^*}^2$ given $\mathcal{P}(A)$ and $\mathcal{P}(N|A)$; however, this form obscures the symmetrical roles of $A$ and $A^*$. The symmetry becomes more apparent if we define the cross-correlation
\begin{equation}
\gamma_{AA^*} \equiv \frac{1}{\sigma_A \sigma_{A^*}} \langle (A-\overline{A}) (A^* - \overline{A^*}) \rangle,
\end{equation}
allowing us to write
\begin{equation}
b^2_{A^*} = \frac{\xi_{A^*}(r)}{\xi_A(r)} = \frac{\sigma_{A^*}^2}{\sigma_A^2} \gamma^2_{AA^*}.
\end{equation}

If $A$ and $A^*$ were perfectly correlated, then $b^2_{A^*}$ would be the ratio of the variances. However, the cross-correlation must be imperfect because one distribution is continuous and the other discrete; at low number densities, this effect becomes most pronounced, because a large range of negative $A$-values (theoretically unbounded below) map to $A^*|_{N=0}$. Thus for low number densities the bias is much smaller than the ratio of the variances, but in the opposite (continuum) limit, $A^*$ approaches $A$ and the bias approaches unity.

\section{Accuracy and Limits}
\label{sec:acc}

\begin{figure}
\leavevmode\epsfxsize=9.5cm\epsfbox{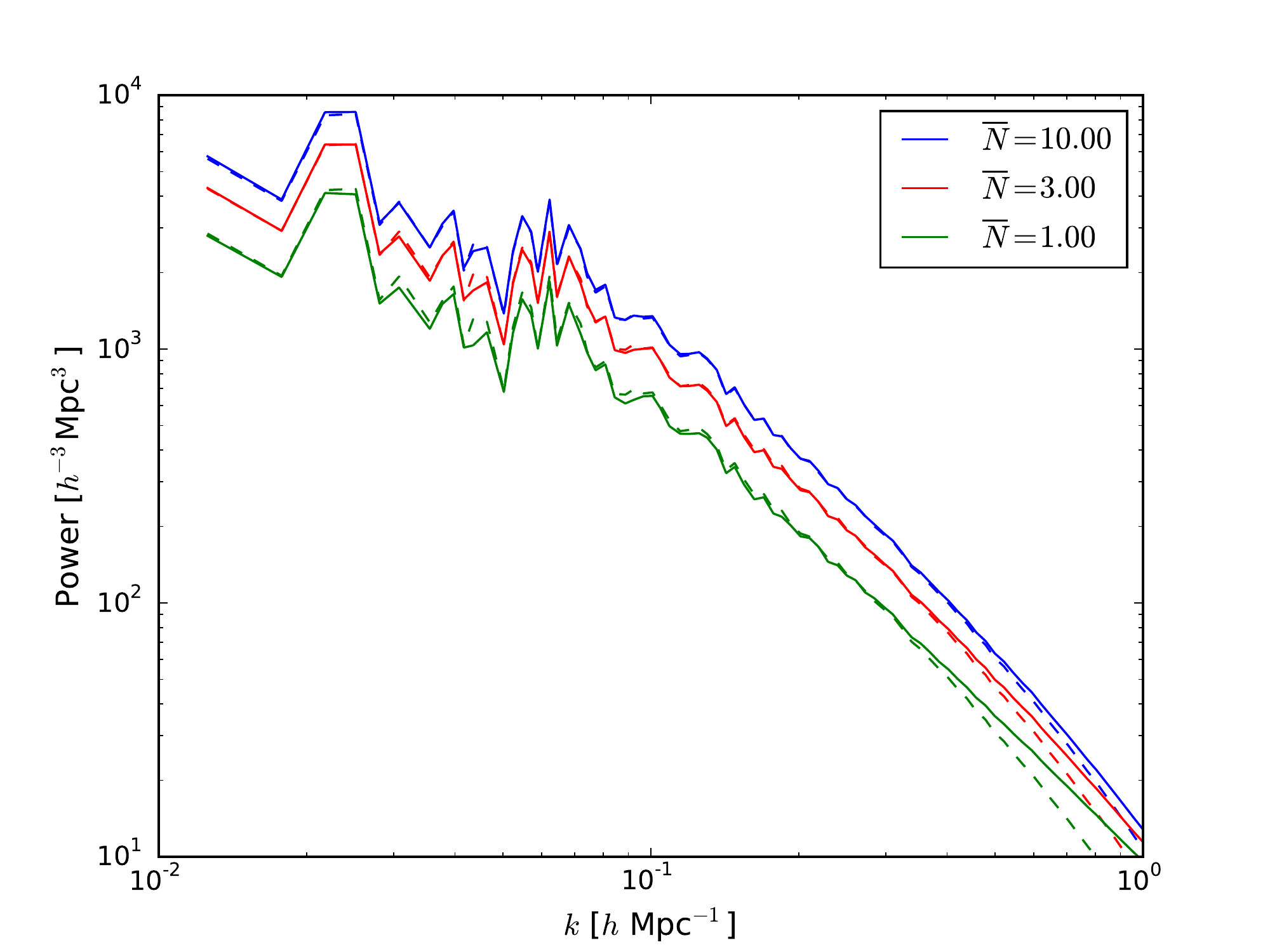}
\caption{Comparison, for three $\overline{N}$-values, of measured spectra $P_{A^*}(k)$ (solid curves) with the spectra $b_{A^*}^2 P_A(k)$ (dashed curves) predicted by Equation~\ref{eq:bAstar}. The colored $P_{A^*}(k)$ curves represent the mean of multiple independent Poisson samplings of the Millennium Simulation dark matter field ($z=0$).} 
\label{fig:spectra}
\end{figure}
To test the accuracy of Equation~\ref{eq:bAstar}, we obtain log dark matter densities $A$ from the Millennium Simulation \citep{MillSim}. From these dark matter fields we generate multiple Poisson realizations for various mean pixel number densities $\overline{N}$. Given the near-GEV distribution of $A$ \citep{predA}, we use Equation~\ref{eq:AstarGEV} to calculate the corresponding $A^*$-field, the power spectrum of which we then measure. Fig.~\ref{fig:spectra} compares $b_{A^*}^2 P_A(k)$ (calculated with Equation~\ref{eq:bAstar}) to the actual power spectrum $P_{A^*}(k)$ for three values of $\overline{N}$; clearly Equation~\ref{eq:bAstar} provides a reasonable estimate of the actual $A^*$-bias as long as $\overline{N}$ is not overly small.

We thus turn to more extensive quantification of the accuracy of this equation; we also compare its prescription with that of \citet{WCS2015}, who give\begin{equation}
b_{A^*}^2 = \left(1 + \frac{1}{\overline{N}\sigma_A^2}\right)^{-2}
\label{eq:WCS}
\end{equation}
for a Poisson-sampled lognormal field. (Strictly speaking, Equation~\ref{eq:WCS} applies to projected two-dimensional data, for which the lognormality assumption is more warranted.) We note that in the limit of high number densities (in particular, when $\overline{N}\sigma_A^2 \gg 1$ so that $A^*$ approaches $A$), the bias in both Equations~\ref{eq:bAstar} and \ref{eq:WCS} approaches unity.

We begin with the matter densities of the Millennium Simulation at three redshifts ($z=0.0$, 1.0, and 2.1). We also specify a variety of (initial) average pixel number densities ranging from 0.01 to 30 objects per pixel; since the Millennium Simulation pixels measure $1.95h^{-1}$ Mpc on each side, these values yield volume number densities from 0.0013$h^3$ to 4.0$h^3$ Mpc$^{-3}$, the lowest value corresponding roughly to the estimated \textit{Euclid} number density \citep{EuclidRedBook}. For each combination of redshift and initial number density, we then generate mock galaxy catalogs by Poisson sampling the dark matter field at the specified $\overline{N}$-values. Next, we rebin each of these catalogs to pixels with side length twice, four times, and eight times that of the original. In this manner we generate a range of mock catalogs with pixel side lengths ranging 1.95 to 15.6$h^{-1}$ Mpc and pixel number densities $\overline{N}$ from 0.01 to $\sim10^4$.

\begin{figure}
\leavevmode\epsfxsize=9.5cm\epsfbox{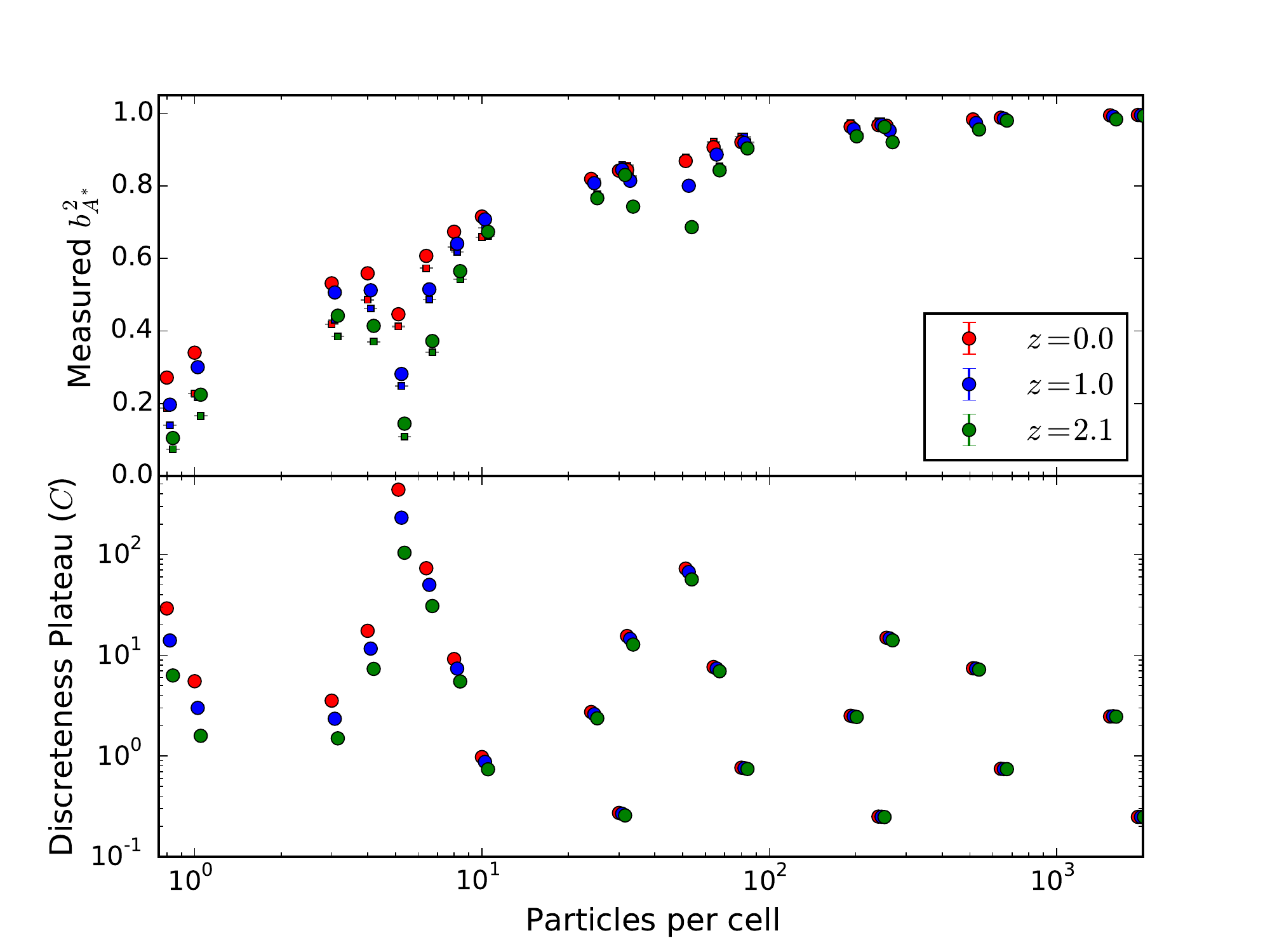}
\caption{Top panel: measured $A^*$-bias values from multiple discrete realizations of the Millennium Simulation dark matter distribution (see text). Circles denote the bias assuming a GEV log matter distribution (Equation~\ref{eq:AstarGEV}); small sqaures denote the bias assuming a lognormal matter distribution (Equation~\ref{eq:Astarln}). Error bars are typically smaller than the size of the markers. Bottom panel: height $C$ of discreteness plateau estimated from mock Poisson surveys, as described in text. For clarity, we apply a small horizontal offset to non-zero redshift markers in both panels.} 
\label{fig:C_and_bias}
\end{figure}

\begin{figure}
\leavevmode\epsfxsize=9.5cm
\epsfbox{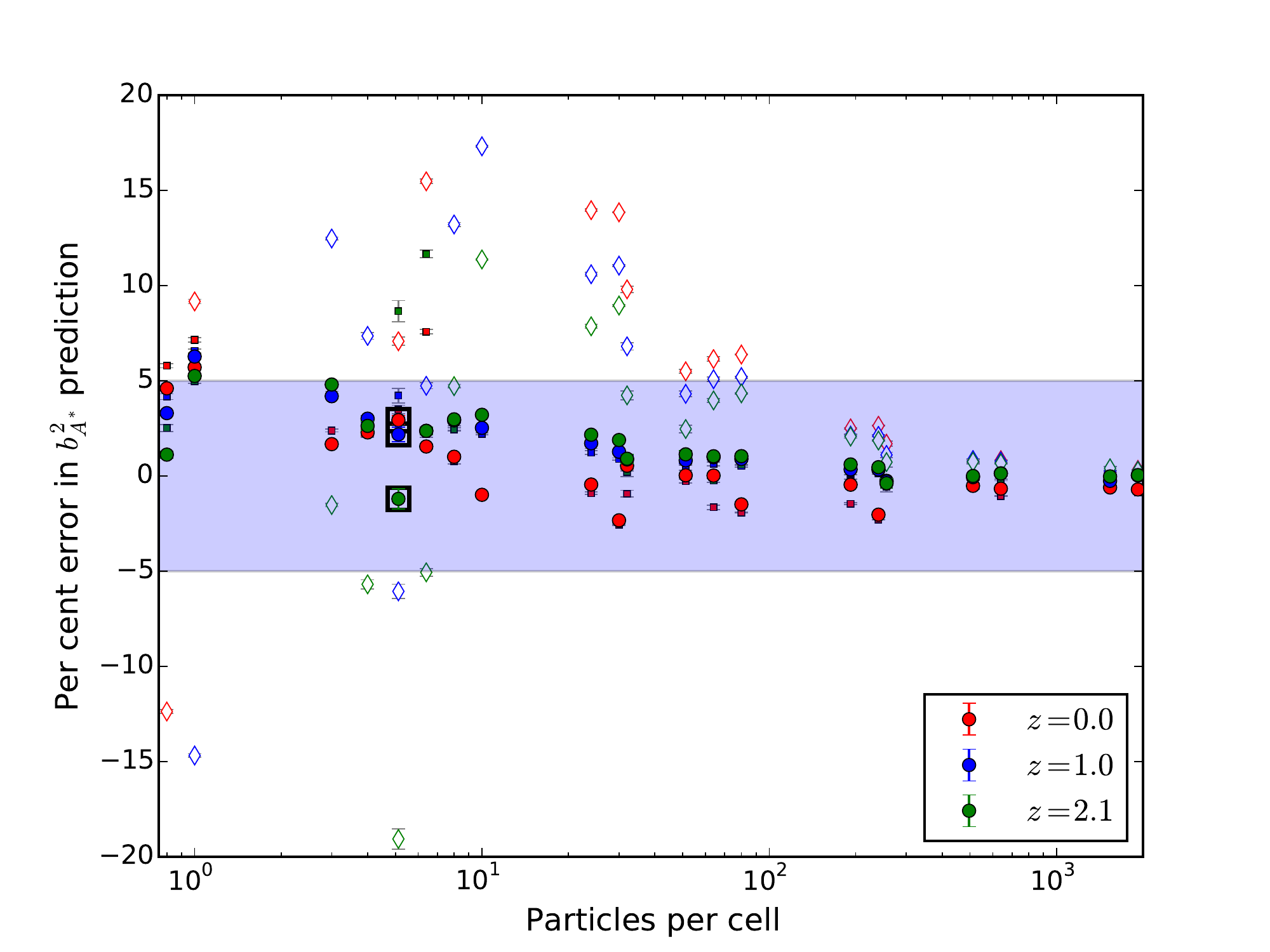}
\caption{Per cent error in two prescriptions for the $A^*$-bias, obtained by measuring the bias from multiple discrete realizations of the Millennium Simulation dark matter distribution (see text). Circles denote the per cent error of Equation~\ref{eq:bAstar}; diamonds denote the per cent error of Equation~\ref{eq:WCS}. Small squares denote the error from using a lognormal distribution (Eqn.~\ref{eq:Astarln}) to define $A^*$. The shaded region highlights the range of less than 5 per cent error; Equation~\ref{eq:bAstar} typically falls within this range for $\overline{N} \ga 1$ per cell. Large black squares highlight symbols corresponding to a \textit{Euclid}-like number density in cubical cells of side length 15.6 $h^{-1}$Mpc. Typical error bars are smaller than the markers.}
\label{fig:Nbar_plot}
\end{figure}

For each catalog we then use Equation~\ref{eq:AstarGEV} to calculate $A^*$ for each pixel and obtain the $A^*$-power spectra to compare with the known $P_A(k)$. To model the discreteness plateau, we use a Monte Carlo approach, populating the mock survey volume with a Poisson distribution having the requisite $\overline{N}$; for this purely Poisson distribution we then calculate $A^*$ and its power spectrum, the (constant) value of which approximates the height $C$ of the plateau. The bias $b_{A^*}^2$ in each mock catalog is then the average value of $(P_{A^*}(k) - C)/P_A(k)$, where we restrict the average to linear wavenumbers ($k < 0.1h$ Mpc$^{-1}$). In this way we attempt to avoid the intermediate- to high-wavenumber effects discussed in Section~\ref{sec:Astar} and shown in Fig.~\ref{fig:Astar} (namely, the discreteness plateau and shape change). Following this procedure, we generate enough mock catalogs (for each redshift and number density) to reduce the uncertainty of the measured bias to less than 0.5 per cent. The resulting measured biases (and constants $C$) appear in Fig.~\ref{fig:C_and_bias}. We also predict the biases using Equation~\ref{eq:bAstar} and the \citet{predA} prescription for the moments of $A$; the results appear in Figure~\ref{fig:Nbar_plot}, which compares the measured biases to those predicted by Equations~\ref{eq:bAstar} and \ref{eq:WCS}. In both Figs.~\ref{fig:C_and_bias} and \ref{fig:Nbar_plot} we also show the impact of assuming a lognormal distribution for $A$, using Equation~\ref{eq:Astarln} to calculate $A^*$.

The prescriptions' accuracy increases as we approach the continuum limit; in this limit $A^* = A$, and both Equations~\ref{eq:bAstar} and \ref{eq:WCS} approach unity. At number densities $\overline{N} \la 100$ particles per pixel, the error in Equation~\ref{eq:WCS} begins to exceed 5 per cent. At the other extreme, the lowest number density we need realistically consider is $\overline{N} \sim 1$, at which point shot noise overwhelms any power spectrum signal. Even down to this number density, Equation~\ref{eq:bAstar} retains its accuracy (in most cases) at the 5 per cent level, although the error rises to 6 per cent for $\overline{N} = 1$ in cells of side length 1.95 $h^{-1}$Mpc. The experiment most closely corresponding to \textit{Euclid} (highlighted by boxes in Fig.~\ref{fig:Nbar_plot}) is the case of volume number density $\overline{n} = 0.0013h^3$ Mpc$^{-3}$, measured in cubical cells of side length 15.6$h^{-1}$ Mpc; the resultant pixel number density is $\overline{N} = 5.1$, with an error in Equation~\ref{eq:bAstar} of less than 3 per cent.

We note that this treatment ignores the issue of galaxy bias; however, \citet{SzapudiPan2004} outline a method for predicting galaxy bias given the underlying dark matter distribution. Fig.~\ref{fig:Nbar_plot} also shows that the error in Equation~\ref{eq:bAstar} depends in part on the underlying one-point statistics. We leave to future work both further refinement of our GEV prescription and its application to galaxy bias.

\section{Conclusion}
\label{sec:concl}
Cosmological surveys probing small scales contain significant information which the standard power spectrum does not access. To retrieve this information, one must instead analyze a sufficient statistic (optimal observable), whose power spectrum and mean together capture virtually all cosmological information in the field.

The log overdensity $A$ is essentially optimal for continuous cosmological fields \citep{CarronSzapudi2013}, but for the discrete fields probed by galaxy surveys, the sufficient statistic is $A^*$ rather than $A$ \citep{CarronSzapudi2014}. One must therefore characterize the power spectrum $P_{A^*}(k)$ to enable maximal utilization of galaxy surveys' information. This power spectrum differs from that of $P_A(k)$ in several respects, the most important of which is the multiplicative bias $b_{A^*}^2$ \citep{WCS2015}.

In this Letter we have characterized this bias for discrete surveys (Equation~\ref{eq:bAstar}) in terms of the underlying log distribution $A$ and the discretization scheme. By generating multiple discrete realizations of the Millennium Simulation dark matter distribution, we have shown the typical error of this prediction to be less than 5 per cent for pixel number densities $\overline{N} \ga 1$ per cell. For survey densities comparable to that of \textit{Euclid}, the error is less than 3 per cent for cubical survey cells of side length $15.6 h^{-1}$ Mpc. This prediction is the most significant portion of the task of characterizing the $A^*$-power spectrum.

Pixels of side length $15.6 h^{-1}$ Mpc correspond to a maximum wavenumber $k \sim 0.2h$ Mpc$^{-1}$, and Figure~\ref{fig:Astar} indicates that $P_{A^*}(k)$ begins to depart from $b_{A^*}^2 P_A(k)$ around this point. Even analysis restricted to this scale represents an 8-fold information gain compared to the linear regime. A hypothetical $L_*$-galaxy survey, on the other hand, might achieve number densities of $0.02$ Mpc$^{-3}$, permitting analysis at much higher wavenumbers. For such a survey, proper treatment of the intermediate- and high-$k$ effects (shape change and discreteness plateau) would be important.

In future work we plan to characterize this high-$k$ discreteness plateau as well as the related shape change; the result will be a complete characterization of the $A^*$-power spectrum. Other work will include a proper accounting for the effects on the log power spectrum of galaxy bias and redshift-space distortion. Preliminary investigation suggests that the log transform renders both of these effects more tractable than they would otherwise be. 

In summary, $A^*$ is a sufficient statistic for galaxy surveys and thus captures all cosmological information inherent in such a survey. To access this information, one must be able to predict the power spectrum $P_{A^*}(k)$ for various sets of cosmological parameters. This Letter provides a key component of this prediction, which will facilitate extraction of maximal information from dense galaxy surveys.

\section*{Acknowledgements}
The Millennium Simulation data bases used in this Letter and the web application providing online access to them were constructed as part of the activities of the German Astrophysical Virtual Observatory (GAVO). IS acknowledges support from National Science Foundation (NSF) award 1616974. 

\bibliography{AStarBias}

\begin{thebibliography}{}
\makeatletter
\relax
\def\mn@urlcharsother{\let\do\@makeother \do\$\do\&\do\#\do\^\do\_\do\%\do\~}
\def\mn@doi{\begingroup\mn@urlcharsother \@ifnextchar [ {\mn@doi@}
  {\mn@doi@[]}}
\def\mn@doi@[#1]#2{\def\@tempa{#1}\ifx\@tempa\@empty \href
  {http://dx.doi.org/#2} {doi:#2}\else \href {http://dx.doi.org/#2} {#1}\fi
  \endgroup}
\def\mn@eprint#1#2{\mn@eprint@#1:#2::\@nil}
\def\mn@eprint@arXiv#1{\href {http://arxiv.org/abs/#1} {{\tt arXiv:#1}}}
\def\mn@eprint@dblp#1{\href {http://dblp.uni-trier.de/rec/bibtex/#1.xml}
  {dblp:#1}}
\def\mn@eprint@#1:#2:#3:#4\@nil{\def\@tempa {#1}\def\@tempb {#2}\def\@tempc
  {#3}\ifx \@tempc \@empty \let \@tempc \@tempb \let \@tempb \@tempa \fi \ifx
  \@tempb \@empty \def\@tempb {arXiv}\fi \@ifundefined
  {mn@eprint@\@tempb}{\@tempb:\@tempc}{\expandafter \expandafter \csname
  mn@eprint@\@tempb\endcsname \expandafter{\@tempc}}}

\bibitem[\protect\citeauthoryear{{Carron} \& {Szapudi}}{{Carron} \&
  {Szapudi}}{2013}]{CarronSzapudi2013}
{Carron} J.,  {Szapudi} I.,  2013, \mn@doi [\mnras] {10.1093/mnras/stt1215},
  \href {http://adsabs.harvard.edu/abs/2013MNRAS.434.2961C} {434, 2961}

\bibitem[\protect\citeauthoryear{{Carron} \& {Szapudi}}{{Carron} \&
  {Szapudi}}{2014}]{CarronSzapudi2014}
{Carron} J.,  {Szapudi} I.,  2014, \mn@doi [\mnras] {10.1093/mnrasl/slt167},
  \href {http://adsabs.harvard.edu/abs/2014MNRAS.439L..11C} {439, L11}

\bibitem[\protect\citeauthoryear{{Carron}, {Wolk}  \& {Szapudi}}{{Carron}
  et~al.}{2015}]{InfoPlat}
{Carron} J.,  {Wolk} M.,   {Szapudi} I.,  2015, \mn@doi [\mnras]
  {10.1093/mnras/stv1595}, \href
  {http://adsabs.harvard.edu/abs/2015MNRAS.453..450C} {453, 450}

\bibitem[\protect\citeauthoryear{{Klypin}, {Prada}, {Betancort-Rijo}  \&
  {Albareti}}{{Klypin} et~al.}{2017}]{Klypin2017}
{Klypin} A.,  {Prada} F.,  {Betancort-Rijo} J.,   {Albareti} F.~D.,  2017,
  preprint, \href {http://adsabs.harvard.edu/abs/2017arXiv170601909K} {}
  (\mn@eprint {arXiv} {1706.01909})

\bibitem[\protect\citeauthoryear{{Laureijs} et~al.}{{Laureijs}
  et~al.}{2011}]{EuclidRedBook}
{Laureijs} R.,  et~al., 2011, preprint, \href
  {http://adsabs.harvard.edu/abs/2011arXiv1110.3193L} {} (\mn@eprint {arXiv}
  {1110.3193})

\bibitem[\protect\citeauthoryear{{Neyrinck}, {Szapudi}  \& {Szalay}}{{Neyrinck}
  et~al.}{2009}]{NSS09}
{Neyrinck} M.~C.,  {Szapudi} I.,   {Szalay} A.~S.,  2009, \mn@doi [\apjl]
  {10.1088/0004-637X/698/2/L90}, \href
  {http://adsabs.harvard.edu/abs/2009ApJ...698L..90N} {698, L90}

\bibitem[\protect\citeauthoryear{{Repp} \& {Szapudi}}{{Repp} \&
  {Szapudi}}{2017a}]{predA}
{Repp} A.,  {Szapudi} I.,  2017a, preprint, \href
  {http://adsabs.harvard.edu/abs/2017arXiv170508015R} {} (\mn@eprint {arXiv}
  {1705.08015})

\bibitem[\protect\citeauthoryear{{Repp} \& {Szapudi}}{{Repp} \&
  {Szapudi}}{2017b}]{Repp2017}
{Repp} A.,  {Szapudi} I.,  2017b, \mn@doi [\mnras] {10.1093/mnrasl/slw178},
  \href {http://adsabs.harvard.edu/abs/2017MNRAS.464L..21R} {464, L21}

\bibitem[\protect\citeauthoryear{{Shin}, {Kim}, {Pichon}, {Jeong}  \&
  {Park}}{{Shin} et~al.}{2017}]{Shin2017}
{Shin} J.,  {Kim} J.,  {Pichon} C.,  {Jeong} D.,   {Park} C.,  2017, preprint,
  \href {http://adsabs.harvard.edu/abs/2017arXiv170506863S} {} (\mn@eprint
  {arXiv} {1705.06863})

\bibitem[\protect\citeauthoryear{{Springel}, {White}, {Jenkins}, {Frenk},
  {Yoshida}, {Gao}, {Navarro}  et~al.}{{Springel} et~al.}{2005}]{MillSim}
{Springel} V.,  {White} S.~D.~M.,  {Jenkins} A.,  {Frenk} C.~S.,  {Yoshida} N.,
   {Gao} L.,  {Navarro} J.,   et~al., 2005, \mn@doi [\nat]
  {10.1038/nature03597}, \href
  {http://adsabs.harvard.edu/abs/2005Natur.435..629S} {435, 629}

\bibitem[\protect\citeauthoryear{{Szapudi} \& {Pan}}{{Szapudi} \&
  {Pan}}{2004}]{SzapudiPan2004}
{Szapudi} I.,  {Pan} J.,  2004, \mn@doi [\apj] {10.1086/380920}, \href
  {http://adsabs.harvard.edu/abs/2004ApJ...602...26S} {602, 26}

\bibitem[\protect\citeauthoryear{{Uhlemann}, {Codis}, {Pichon}, {Bernardeau}
  \& {Reimberg}}{{Uhlemann} et~al.}{2016}]{Uhlemann2016}
{Uhlemann} C.,  {Codis} S.,  {Pichon} C.,  {Bernardeau} F.,   {Reimberg} P.,
  2016, \mn@doi [\mnras] {10.1093/mnras/stw1074}, \href
  {http://adsabs.harvard.edu/abs/2016MNRAS.460.1529U} {460, 1529}

\bibitem[\protect\citeauthoryear{{Wolk}, {Carron}  \& {Szapudi}}{{Wolk}
  et~al.}{2015}]{WCS2015}
{Wolk} M.,  {Carron} J.,   {Szapudi} I.,  2015, \mn@doi [\mnras]
  {10.1093/mnras/stv1891}, \href
  {http://adsabs.harvard.edu/abs/2015MNRAS.454..560W} {454, 560}

\makeatother
\end{thebibliography}
\label{lastpage}
\end{document}